# CORONAS-F OBSERVATION OF GAMMA-RAY EMISSION FROM THE SOLAR FLARE ON 2003 OCTOBER 29

Victoria G. Kurt[1], Boris Yu. Yushkov[1], Vladimir I. Galkin[1,2], Karel Kudela[3,4*],
and Larisa K. Kashapova[5]

[1] *Skobeltsyn Institute of Nuclear Physics, Lomonosov Moscow State University, Moscow, 119991, Russia*

[2] *Faculty of Physics, Lomonosov Moscow State University, Moscow, 119991, Russia*

[3] *Institute of Experimental Physics, Slovak Academy of Sciences, Košice, 04001, Slovakia*

[4] *Nuclear Physics Institute, Czech Academy of Sciences, Řež, Czech Republic*

[5] *Institute of Solar-Terrestrial Physics, Russian Academy of Sciences, Irkutsk, 664033, Russia*

## ABSTRACT

Appreciable hard X-ray (HXR) and gamma-ray emissions in the 0.04–150 MeV energy range associated with the 2003 October 29 solar flare (X10/3B) were observed at 20:38–20:58 UT by the SONG instrument onboard the CORONAS-F mission. To restore flare gamma-ray spectra we fitted the SONG energy loss spectra with a three-component model of the incident spectrum: (1) a power law in energy, assumed to be due to electron bremsstrahlung; (2) a broad continuum produced by prompt nuclear de-excitation gamma-lines; and (3) a broad gamma-line generated from pion-decay. We also restored spectra from the *RHESSI* data, compared them with the SONG spectra and found a reasonable agreement between these spectra in the 0.1–10 MeV energy range. The pion-decay emission was observed from 20:44:20 UT and had its maximum at 20:48–20:51 UT. The power-law spectral index of accelerated protons estimated from the ratio between intensities of different components of gamma rays changed with time. The hardest spectrum with a power-law index $S$ = -3.5 to -3.6 was observed at 20:48–20:51 UT. Time histories of the pion-decay emission and proton spectrum were compared with changes of the locations of flare energy release as shown by RHESSI hard X-ray images and remote Hα brightening. An apparent temporal correlation between processes of particle acceleration and restructuring of flare magnetic field was found. In particular, the protons were accelerated to sub-relativistic energies after radical change of the character of footpoint motion from a converging motion to a separation motion.

* corresponding author kkudela@saske.sk

# 1. INTRODUCTION

The processes of particle acceleration during solar flares can be studied through observations of hard X-ray (HXR) and gamma-ray emissions. This is especially the case with the acceleration of relativistic electrons and high-energy protons. Electrons produce a bremsstrahlung continuum with energies up to those of electrons themselves (Miller & Ramaty 1989). The potential value of gamma-rays as a probe of energetic ions accelerated in solar flares was pointed out and the expected fluxes were calculated in detail by Lingenfelter & Ramaty (1967) and Lingenfelter (1969). Protons with energies of tens of MeV excite nuclei of ambient matter, which emit gamma-lines in the 0.5–12 MeV energy range (Ramaty & Murphy 1987; Murphy et al. 2007, 2009). If the protons gain energies above 300 MeV then they can produce neutral and charged pions through interactions with the solar atmospheric material. These pions in turn decay and generate gamma-ray emission with a specific spectrum, namely a broad plateau in the 30–150 MeV energy range (e.g., Murphy & Ramaty 1984; Ramaty & Murphy 1987; Murphy et al. 1987). The pion-decay emission was observed in several solar flares starting from 1982 June 3 (Forrest et al. 1985, 1986). Brief reports of these observations can be found in Chupp et al. (2003), and Chupp & Ryan (2009). Results on solar gamma rays and neutrons have been summarized until 2011 along with the open questions by Vilmer et al. (2011). High-energy gamma-ray emission from several major solar flares was observed by the *Solar Neutrons and Gamma-rays* (SONG) detector onboard the *Complex Orbital Observations of the Active Sun* (CORONAS-F) mission, namely on 2001 August 25, 2003 October 28, 2003 November 4, and 2005 January 20, and the pion-decay component was reliably measured in these events (Kurt et al. 2010, 2013; Kuznetsov et al. 2011). During the current solar activity cycle, high-energy gamma-rays (above 100 MeV) were observed by the *Fermi* Large Area Telescope in and after several solar flares (Ackermann et al. 2012, 2014; Ajello et al. 2014; Pesce-Rollins et al. 2015). The list of events with observations of high-energy gamma-rays during the flare's impulsive phase remains very short, so the increase in this number is very important.

Due to the experience obtained through the analysis of events observed by SONG, we improved our methods and became capable of analysis flares with less intense pion-decay gamma-emission. We report here on a comprehensive study of HXR and gamma-ray emissions from the 2003 October 29 X10 flare (W05 S18) observed by CORONAS-F/SONG in the 0.04–150 MeV energy range. These observations lasted for 20 minutes and covered entirely the flare impulsive phase, providing an opportunity to track the temporal evolution of the gamma-emission spectrum including the pion-decay component. This flare was one of a series of flares originated from NOAA AR 10486.

The flare had a large amount of free magnetic energy, ~ 6·10$^{33}$ ergs, which is consistent with the very high level of activity observed in this active region (see, e.g., Veselovsky et al. 2004; Metcalf et al. 2005). Emissions associated with the SONG event were observed from meterwave, microwave, optical, ultraviolet and solar energetic particles as a ground-level enhancement.

This flare was also observed by the *Ramaty High Energy Solar Spectroscopic Imager* (*RHESSI*). Hurford et al. (2006) carried out the *RHESSI* imaging in the 2.223 MeV gamma line and compared the imaged and spatially integrated fluences in this line. The authors showed that most, if not all, of the emission was confined to the compact flare region. They concluded that gamma-ray-producing ions with energies of 10–100 MeV "appear to be accelerated by the flare process". Most available publications based on the *RHESSI* data deal with the photon energy range from 6 to 150 keV. The motions of significant HXR footpoints (FPs) were analyzed by Krucker et al. (2005), Ji et al. (2008), Des Jardins et al. (2009), Liu et al. (2009), Xu et al. (2010), and Yang et al. (2011). These publications provided us an opportunity to compare the change of high-energy emission spectra with conspicuous changes in the structure of the flare region.

In Section 2, we present observations of the flare by the SONG and other experiments and examine main stages of the flare development. In Section 3, we describe routines of restoration of HXR and gamma-emission spectra from the SONG and *RHESSI* data and examine these spectra. In Section 4 we study evolution of separate components of HXR and gamma-emission and estimate spectra of high-energy flare-accelerated particles generating observed HXR and gamma-emission. In Section 5, we compare an evolution of populations of high-energy flare-accelerated particles with an evolution of the flare magnetic field structure obtained from *RHESSI* and Hα image observations. Conclusions of the paper are given in Section 6.

## 2. OBSERVATIONS

The SONG instrument had a CsI crystal with a diameter of 20 cm and a height of 10 cm as the main detecting element. The crystal was surrounded by an anticoincidence plastic scintillator shield to reject signals from charged particles. Count rates caused by HXR and gamma-rays were recorded in 12 channels formed by pulse-height discriminators. A brief description of the SONG was presented by Kuznetsov et al. (2011); this paper also contains a description of the in-flight-calibration procedure and the background subtraction routine.

The flare on 2003 October 29 was observed by CORONAS-F under favorable conditions: near the equator in the eastern hemisphere, where background count rates were low. Figure 1 shows the response of the SONG detector to the flare HXR and gamma-ray emissions over the range of energy losses from 0.04 to 150 MeV. The net count rates due to the flare (4-s time bins) were obtained by subtracting the background count rates. The count rates above 6 MeV were then accumulated in 40-s bins to improve poor statistics. Figure 1 also shows the time profiles of the time derivative of the *Geostationary Operational Environmental Satellite* (*GOES*) soft X-ray (SXR) at 1-8 Å, $dI_{SXR}/dt$, the mean thermal plasma temperature derived from the GOES two-channle emission, the microwave emnission at 15.6 GHz, and the *RHESSI* HXR count rates, respectively.

Figure 1 shows that the time profiles of count rates in different energy channels differ noticeably. Three broad structured enhancements are clearly seen at the energies below 10 MeV. They coincide with major enhancements in the *RHESSI* 80-120 keV and microwave light curves. A complicated structure of first and second enhancements was studied in details by Kurt et al. (2015). These enhancements were higher than the third one at energies below 750 keV. All enhancements leveled off at 0.75–10.5 MeV (panels *d-g*). The start times of the second and the third enhancements are marked as $t_1$ and $t_2$, respectively. These times, $t_1$ and $t_2$, are not exact boundaries but more likely they specify time intervals of transitions between consecutive energy releases. Note that $t_1$ divides phases I and II which were introduced by Liu et al. (2009) based on HXR FP motions. Brief review of image observations is presented in Section 5. It is observed that, after t2, the decrease of the mean plasma temperature deduced from the GOES SXR emissions has slowed down, suggesting that another episode of energy release may be taking place generating more hot plasmas.

The flare emission in the 10–22.5 MeV channel began to arise after 20:44:30 UT (panel *g*). At the energies above 22.5 MeV, a statistically significant flare emission was observed only after 20:46 (panels *h–l*). Its intensity peaked at 20:49–20:51 UT and decreased afterward.

A visible existence of the second and the third enhancements permits us to separate their contributions in the count rates at energies below 10.5 MeV after 20:47 UT. We assumed that an intensity of the second enhancement decayed exponentially from ~ 20:46 UT (see Figure 2) and found the best fits of decay constants $\tau(E)$ in each channel. Thereafter, we subtracted estimated values of count rates from recorded ones and found in this way the initial profiles of the third enhancement that started at 20:47:30 UT ±30 s. Note that there was a statistically significant increase in the 6-10 MeV channel at this time.

## 3. EVOLUTION OF THE FLARE EMISSION SPECTRUM

### 3.1. Restoration of an incident HXR and gamma-ray spectra

The main feature of the 2003 October 29 flare observed by SONG was an appreciable flux of high-energy gamma-rays. Certain of the recorded count spectra were characterized by a wide plateau at 20–150 MeV. Since the classical papers of Forrest et al. (1985, 1986), it is commonly accepted that this spectral feature can be explained by generation and decay of neutral and charged pions (Ramaty & Murphy 1987; Murphy et al. 1987). We also follow this conventional paradigm.

Measurements provided with SONG allow us to restore spectra of HXR and gamma-ray emission from solar flares. To fit the spectra from this flare we used a three-component model consisting of: (1) the bremsstrahlung from primary accelerated electrons, (2) the broad continuum formed by prompt γ-lines from nuclear de-excitation, and (3) the broad continuum produced by pion decay. The bremsstrahlung spectrum is described as a power law with roll-over at high energies:

$$F(E) = I_0 E^{-\gamma} \exp(-\frac{E}{E_0}) \qquad (1)$$

The energy resolution of CsI crystals and broad energy channels of SONG did not allow us to separate discrete narrow gamma-lines. That is why we consider integrated deposit from these lines into the SONG count rate in the 0.5–12 MeV energy range and use a broad continuum produced by these lines in our fitting. The shape of this continuum was obtained from Figure 20 of Murphy et al. (2009). The pion-decay continuum mainly consists of a wide gamma line from neutral-pion decay peaking at 67 MeV and bremsstrahlung from secondary positrons. This spectrum component was calculated by R. Murphy and kindly provided to us (private communication). In general, it is similar to a typical shape of this continuum (see, e.g., Figure 18 in Murphy et al. 1987).

The response matrices of the SONG detector were simulated separately for the spectrum components considered above using the GEANT 3.21 program. Then we fit recorded count spectra, find weights of different components and reconstruct incident spectra of HXR and the gamma emission. Figure 3a presents a comparison of the recorded and fitted count spectra during the maximum of the high-energy gamma-rays. Figure 3b presents the incident photon spectrum and the same count spectrum recalculated to the photon one using the fitted detector response. This Figure demonstrates that the total spectrum below 0.75 MeV is caused entirely by the bremsstrahlung, between 0.75 and 10.5 MeV mainly by the γ-lines, and above 10.5 MeV by the pion-decay emission.

We simulated the SONG response matrices for various shapes of spectrum of the pion-decay component, which were calculated for different values of α/proton ratio (from 0.1 to 0.5) and indices of the parent proton power-law spectrum - from 2.0 to 4.0 (R. Murphy, private communication). We found that these variations change the fitted weight of the pion-decay component no more than a few percent, which is negligible in our case. Hereafter we use the spectrum of the pion-decay component calculated for the proton-spectrum index equal to 3.0 and the α/p ratio equal to 0.5.

In principle, a similar gamma-ray spectrum with a plateau at high energies could be caused by some additional population of electrons accelerated to ultra-relativistic energies with a very hard spectrum in the energy range of 10–200 MeV. Such a possibility was discussed, for example, with regard to the flare of 2010 June 12 measured by the *Fermi* LAT (Ackermann et al. 2012). However, the authors of the latter paper themselves preferred the pion-decay origin of the high-energy gamma-emission. Furthermore, strong arguments can be brought against the high-energy electron hypothesis. Acceleration mechanisms, which could create an electron spectrum with an almost flat high-energy tail, are unknown. These hypothetical high-energy electrons cannot also be a simple extension of the real population of low-energy electrons, which produced the observed bremsstrahlung, due to an evident difference in the power-law indices.

Another argument in our case may be the following. The flare was located at the solar disk center (W05). An opening cone of bremsstrahlung is of the order of $m_ec^2/E_e$, which is about 0.5° for 100 MeV electrons. Multiple scattering of these electrons is also very small - about 10° in one radiation length, corresponding to 63 g·cm$^{-2}$ of hydrogen. This means that such electrons lose their energy by Coulomb collisions faster than they scatter by a large angle. These two factors could lead to a significant decrease in intensity of radiation of ultra-relativistic electrons. Calculation of bremsstrahlung emissivity taking into account synchrotron energy losses by Dermer & Ramaty (1986), Ramaty et al. (1994), and Park et al. (1997) found considerable steepening of the gamma-ray spectrum above tens of MeV. For example, Dermer & Ramaty (1986) estimated this steepening at 100 MeV as $\Delta\gamma = 1.8$ for the bremsstrahlung perpendicular to the electron beam direction.

High-energy photons can be also produced by electrons with similar energies or protons with energies above 300 MeV through their interactions with the Earth's atmosphere. However, the most intense pulse of the pion-decay emission had 3-min duration thus it could not be a terrestrial gamma flash or a similar transient. Electrons with similar energies are absent in the magnetosphere in considerable amount. Moreover, the satellite was located in this time (47:55-51:04 UT) at 169°E between 10° N and 3° S. Its altitude was 415 km, thus a visible range of the Earth's atmosphere was limited by 20°. This range corresponds to cut-off rigidities above 10 GV that is more than maximum energies recorded by the neutron monitor network during Ground Level Enhancement (GLE) 66 accompanied this flare. Besides, this GLE began later, after 21:00 UT. Therefore, any terrestrial origin of the observed enhancement of the high-energy gamma emission must be excluded. Basing on these considerations, we believe that the high-energy plateau visible in the spectra restored from the SONG data is really caused by the solar pion-decay emission.

Besides spectra obtained from the SONG data we also studied spectra obtained from the *RHESSI* data. These spectra were obtained using the package for *RHESSI* data processing and the SPectrum EXecutive (SPEX) software (Schwartz et al. 2002). The *RHESSI* spacecraft crossed a 'spur' of the South Magnetic Anomaly from 20:46:30 to 20:55 UT. That is why there was a contribution to count rates of the front detectors (see Figure 1 from Liu et al. 2009) and especially of the rear detectors by high-energy protons (see Hurford et al. 2006). In order to avoid this problem and compare the low-energy spectra obtained by *RHESSI* and SONG we applied the method of the imaging spectroscopy to obtain spectra from the front detectors (energies below 300 keV). This method allows one to get a net flux from the flare emission that is free of charged particle contribution (see Liu et al. 2009). We used the 9th detector as the thickest and most sensitive one and reconstructed images using PIXON algorithms (Puetter & Pina 1994; Metcalf et al. 1996) for energies above 30 keV. The low-energy limit was chosen so as to escape influence of thermal spectrum component. For each time interval, we integrate the photon flux at given energy bands in an area determined from the largest HXR source.

The spectra above 300 keV are the averaged ones obtained from all *RHESSI* rear detectors. The model of the gamma-ray spectrum for fitting was kindly provided by G. Share (private communication). It consists of functions describing the bremsstrahlung spectrum and emission of the lines. The bremsstrahlung spectrum is described by a power-law function for energies below 1 MeV, denoted as PL1 following Ackermann et al. (2012), and a power law with roll-over for higher energies (PL2) described by Eq. (1). The nuclear de-excitation lines and the continua are represented by a template based on a detailed study of nuclear gamma-ray production from accelerated-particle interactions with elements found in the solar atmosphere (Murphy et al. 2009). The emission of 511 keV line with the positron continuum was described by a special function. The line at 2.223 MeV was fitted by the symmetric Gaussian function. Figure 4a presents an example of the fitted spectrum. A good agreement of spectra deduced from both front and rear *RHESSI* detectors is seen.

As stated above, the SONG detector could not separate discrete narrow gamma-lines including the well-known line at 2.223 MeV. To find properly a real flux of prompt nuclear gamma-lines measured by SONG we calculated a possible contribution of the neutron-capture line into the SONG count rates in the 0.75–6 MeV channels using the simulated SONG response to this line and its intensity measured by *RHESSI*. Then we subtracted this calculated contribution from recorded count rates and fitted resulted count spectra with the same three-component model. This procedure leads to an expected decrease of the gamma-line flux together with an increase of the pion-decay emission flux (see Figure 4).

Figure 4 demonstrates a reasonable consistency of the spectra obtained from the SONG and *RHESSI* data. The total gamma-line flux measured by SONG and averaged over the range of 4-7 MeV which is crucial for a further analysis, was about 1.6 times higher than the *RHESSI* one. This difference is caused by contribution from the PL2 component into the *RHESSI* flux. The spectral index of the latter component was equal to 0.3 in the case under consideration and ranged from 0.3 to 1.2 during the flare. This calls into question a physical meaning of this component. Note, that the total flux in this energy range measured by *RHESSI* is equal in general to the SONG flux.

In principle, it is possible to use a variety of trial spectrum models to fit observed data. If these models describe realistically an incident spectrum, and responses of detectors were calculated correctly, then incident photon spectra restored from different experiments for the same time intervals have to be close to each other with a reasonable accuracy.

### 3.2. Evolution of the flare-emission spectrum

For further analysis the flare was divided into thirteen successive time intervals according to visible times of count rate changing. These intervals are listed in Table 1. Duration of each interval provided count numbers required to be statistically significant, even in high-energy channels. Background subtracted count-rate spectra were accumulated during these intervals and fitted as was described above. Parameters of spectra fitted from the SONG data and fluxes at fixed energies are presented in Table 1; $\gamma$ and $E_0$ describe the bremsstrahlung spectrum according to Eq. (1). The uncertainty of $\gamma$ determination is ±0.1, of $E_0$ is ±2 MeV. Accuracy of fluxes at 100 keV is 10%. Accuracy of fluxes at 5 MeV was estimated as 30% taking into account the uncertainty of contribution of the 2.223 MeV line.

Figure 5 shows emission spectra derived from the SONG as well as the *RHESSI* measurements. This Figure demonstrates a good agreement between these spectra in the common energy range before 20:46:35 and after 20:55:00 UT despite the difference in the models describing the bremsstrahlung spectrum. The *RHESSI* spectra obtained from 20:46:30 to 20:55 UT exceed the SONG ones most likely due to a contamination by charged particles. At 20:48:48–20:52 UT these spectra contain only PL1, PL2, and 2.2 MeV components without the gamma-line component that can be caused by strong contamination.

Figure 5 demonstrates essential changes in HXR and gamma-ray spectra during the flare course. The most important experimental fact derived from the SONG data is a significant flare gamma-emission above 20 MeV which we have attributed to be caused by the pion-decay process. This emission appeared at 20:44–20:45 UT, i.e., close to $t_1$. Its significant jump was observed near 20:48 UT, i.e., close to $t_2$.

# 4. TIME HISTORY OF FLARE-EMISSION COMPONENTS AND EVOLUTION OF POPULATIONS OF FLARE-ACCELERATED PARTICLES

Our study is based on observations of HXR and gamma-ray emissions which are produced in interactions of flare-accelerated particles with the solar atmosphere material. These emissions except the neutron-capture line are generated immediately after interactions. Therefore observed time history of these emissions will be essentially identical to the time history of interactions of different components of flare-accelerated particles. One must bear in mind, however, that the emission versus time is a complicated superposition of "acceleration–transport–energy loss" processes which takes a certain amount of time. In the present study we ignore this delay and suppose that observed time profiles of HXR and gamma-ray emissions correspond to the time history of particle acceleration. The 2.223 MeV line is delayed relative to appearance of accelerated ions and neutrons due to the neutron thermalization and capture times. This delay varies from one flare to another and ranges from 80 up to 120 s (Prince et al. 1983; Trottet et al. 1993; Dunphy et al. 1999; Murphy et al. 2007).

The three-component spectrum model used for SONG data fitting provides direct estimates of different component intensities that permits us to study time evolution of these components separately (see Figure 6 and Table 2). The bremsstrahlung total flux, $F_{br}$, was calculated at energies >100 keV. An accuracy of this flux is 10%. The bremsstrahlung spectrum power-law index $\gamma$ is also presented in Figure 6a. The flux of the prompt gamma-lines, $F_{4–7}$, was calculated over the 4–7 MeV range, an accuracy of this flux was estimated as 30%. The pion-decay emission flux, $F_\pi$, calculated over the 30–100 MeV range, is presented in Figure 6d in comparison with the SONG count rates at 40–150 MeV. Note that the profile of the pion-decay emission did not match profiles of HXR and prompt nuclear gamma-lines.

Figure 6 contains also fluxes obtained from the *RHESSI* spectra. The gamma-line flux, $F_{4–7}$, was again calculated over the 4–7 MeV range. Fluxes of prompt gamma lines measured by SONG and *RHESSI* were proportional (see panel *b*), the difference between them was explained above.

The profile of the bremsstrahlung with the energy >100 keV produced by primary accelerated electrons, was almost synchronized with the profile of the prompt gamma-lines (see Figures 6a,b) which are a signature of interactions of ions with energies of 5–20 MeV. Both these ions and electrons were accelerated at the very beginning of the flare impulsive phase (see Tables 1 and 2) implying that proton and relativistic electron acceleration is a common property of the impulsive energy release process in solar flares (Forrest & Chupp 1983). The electron spectrum power-law index $\gamma_e$, derived from the bremsstrahlung spectrum in the thick target model, ranged from 5.4 at the impulsive phase onset to ~4–4.2 near its end.

The hardness of proton spectra can be studied through comparison of fluxes of prompt nuclear gamma-lines, the neutron-capture line, and the pion-decay emission because these components are produced by particles belonging to different energy ranges (Ramaty & Murphy 1987). The prompt gamma-lines are produced by ions with energies of 5–30 MeV, whereas the 2.223 MeV line is predominantly caused by 10–60 MeV ions. Observation of the pion-decay emission undeniably proves that the ions were accelerated to energies at least above the pion-production threshold (~300 MeV). Analysis of evolution of these fluxes allows us to study changing of the proton spectrum during the course of the flare.

Figure 6 clearly demonstrates that this spectrum was indeed changing. Two broad peaks in the prompt gamma-lines were equal in amplitude whereas the second peak of the 2.223 MeV gamma line was distinctly higher that the first peak of this line. This implies that the proportion of high-energy (10–60 MeV) ions increased after $t_2$. To scrutinize this fact we analyzed once again evolution of the prompt gamma-lines and the neutron-capture line using 40-s intervals. We restored spectra from the *RHESSI* and found an intensity of the 2.223 MeV gamma line. Then we subtracted a contribution of this line from the SONG count rates and restored flare-emission spectra as it was described above. The time profiles of fluxes of the 2.223 MeV and prompt gamma-line obtained with SONG are presented in Figures 7a,b.

The neutron-capture line at 2.223 MeV was observed within 30-60 s after the onset of impulsive energy release. The time history of this line is related with time history of accelerated particles in a complicated way due to stochastic character of neutron thermalization process. Following Prince et al. (1983),

$$F_{2.2}(t) = \int_{-\infty}^{t} S(t_1) R(t, t_1) \, dt_1, \qquad (2)$$

where $F_{2.2}(t)$ is observed flux of the 2.223 MeV line, $S(t)$ is a neutron production history, taken to be a time history of prompt gamma lines, $F_{4-7}(t)$, and $R(t,t_1)$ – function giving the 2.223 MeV photon yield at time $t$ due to neutrons produced at time $t_1$. This function depends implicitly on a shape of an ion spectrum. In the simplest case when the shape of the ion spectrum does not vary during the flare, we can derive

$$F_{2.2,j} \propto \sum_{i=1}^{j} F_{4-7,i} \exp(-\frac{t_j - t_i}{\tau}), \qquad (3)$$

where $\tau$ is a time constant of an exponential fall-off of the neutron-capture line. For typical flare parameters, this constant is about 100 s.

We simulated an expected time history of the neutron-capture line using Eq. (3) with $\tau = 100$ s, and the prompt gamma-line fluxes, obtained with SONG in 40 s intervals (see Figure 7a), as $F_{4-7,i}$. To compare observed and simulated time profiles of the neutron-capture line we equated magnitudes of these profiles near 20:45 UT (see Figure 7b). A general similarity of observed and simulated profiles is clearly seen but the observed values exceeded simulated ones after 20:48 UT. The residuals between them are also shown in the Figure. Although the errors of the residuals are large enough, this excess is statistically significant because it was visible during ~8 min. An existence of this additional flux of the 2.2 MeV line proves that the flare-accelerated ion spectrum became harder and the proportion of ions with energies above 30 MeV increased after 20:48 UT.

Next we investigate evolution of accelerated ion spectrum. First we will study a low-energy part of an ion spectrum. The well-known and commonly used method of estimating proton spectrum hardness is based on comparison of fluxes of the neutron-capture gamma-line 2.223 MeV and the de-excitation gamma-lines (Hua & Lingenfelter 1987). The ratio of these fluxes, $F_{2.2}/F_{4-7}$, depends on the shape of the particle energy spectra in the 5−60 MeV range. In order to avoid complexities in calculation of $R(t,t_1)$, traditionally, fluxes of neutron-capture and prompt gamma-lines are integrated over the entire duration of the flare or for an enough long time during the flare (see, e.g., Dunphy et al. 1999; Rieger 1996; and Kiener et al. 2006).

In the case being considered we also first calculated the fluences of the neutron-capture and prompt gamma-lines and their ratios in three selected intervals. Then we estimated the power-law indices $S$ using Figure 14 in Hua & Lingenfelter (1987) considering the flare location (W05). Values of $S$ obtained in this way are presented in Figure 7b. The hardest spectrum was found between 20:51−20:56 UT. Next we study the ion spectrum evolution using 40 s data. In order to avoid calculation of $R(t,t_1)$, we computed values of $S$ using various values (40−120 s) of delay of the neutron-capture line relative to the de-excitation gamma-lines as well as various durations of accumulation intervals of the neutron-capture line. Then we averaged values of $S$ obtained in this way. These values coincide well with those obtained for long time intervals (see Figure 7c).

Further we study a high-energy part of an ion spectrum. The ratio $R$ of the pion-decay emission and gamma-lines intensities, $F_\pi/F_{4-7}$, depends strongly on the shape of the accelerated-particle spectrum. The expected $R$ value can be calculated under certain assumptions about the spectrum and spatial distribution of protons as well as about the chemical abundance of the solar atmosphere, and vice versa: the proton spectrum shape can be estimated from this ratio obtained experimentally. Such estimates have been performed, e.g., for the flares on 1982 June 3 (Ramaty & Murphy 1987; Murphy et al. 1987), 1991 June 11 (Dunphy et al. 1999), 2003 October 28 (Kuznetsov et al. 2011), and 2010 June 12 (Ackermann et al. 2012).

We compared the $F_\pi/F_{4-7}$ values obtained from the SONG measurements with theoretical ones. We used Figure 4 from Dunphy et al. (1999), which presents this ratio versus the power-law index $S$, calculated for the downward isotropic proton distribution. Although this ratio was obtained under specific conditions, we believe that these dependencies can also be applied to other events especially for estimates of spectrum variations during the course of a flare. Values of $S$ found in this way are shown in Table 2 and Figure 7d. In the Figure we present spectral indices calculated both for 'long' and 'short' time intervals. Averaged values of $S$ obtained from the $F_{2.2}/F_{4-7}$ ratio in 40 s intervals are also presented in the Figure 7d.

We traced evolution of spectrum of flare-accelerated particles during the flare course. Three time intervals can be clearly seen. A power-law index obtained from the $F_{2.2}/F_{4-7}$ ratio was practically constant (4.2) till 20:46 UT. The pion-decay emission was negligible during this time. Then the ion spectrum became harder. Spectral indices obtained both from $F_\pi/F_{4-7}$ and $F_{2.2}/F_{4-7}$ ratios coincided taking into account their errors. Note that we use the same values of the $F_{4-7}$ flux in both ratios. Therefore the ion spectrum during this time can be described by an unified unbroken power law through wide energy range - from about 5 MeV to several hundred MeV, i.e., above the pion-production threshold. Thus the high-energy gamma emission observed by SONG most likely had the pion-decay origin. The hardest spectrum ($S$ equal to 3.5–3.6) was at 20:49–20:51 UT simultaneously with the maximum intensity of the pion-decay emission. After 20:52 UT the intensity of this emission decreased while intensities of the neutron-capture and prompt gamma-lines increased. The spectral index obtained from the $F_{2.2}/F_{4-7}$ ratio remained hard. This means that the spectrum shape changed and became more complicated than a simple power law. A decrease of the pion-decay emission indicates a spectrum break near 60-100 MeV.

# 5. TEMPORAL RELATIONSHIP BETWEEN ACCELERATION OF HIGH-ENERGY PROTONS AND CHANGES OF STRUCTURE OF FLARE MAGNETIC FIELD

In this section, we compare an evolution of the population of high-energy flare-accelerated particles deduced from the SONG data with an evolution of the flare magnetic field structure inferred from *RHESSI* and Hα observations. The *RHESSI* imaging observations at energies of 6–150 keV were analyzed independently by Ji et al. (2008), Liu et al. (2009), and Xu et al. (2010). These papers provided a detailed picture of motion of two brightest conjugated HXR FPs. Ji et al. (2008) divided the flare process into two phases with a short transition interval between them near $t_1$. During Phase I (from 20:39:20 to ~ 20:44 UT, i.e. till $t_1$) there was converging motion of FPs. During Phase II (from 20:44) FPs moved away from each other. Figure 8a presents the distance between FPs observed in the 70-150 keV range. Ji et al. (2008) named Phase I as a "sigmoid" one and Phase II as an "arcade" one assuming essential changes in the structure of the flare magnetic field. Xu et al. (2010) demonstrated a visible change of the spatial structure of HXR emission between $t_1$ and $t_2$.

Two phases of the conjugated FP motion were also observed by Liu et al. (2009). During Phase I (from 20:40:40 to 20:44 UT) FPs generally moved towards each other in a direction essentially parallel to the magnetic field neutral line (NL). During Phase II, after 20:44 UT FPs moved away from each other mainly perpendicular to the NL. Figure 8b presents the distance between FPs and its components. Figure 8c presents the shear angle - the angle between the line connecting the two conjugate FPs and the normal to the NL. This angle decreased fast during Phase I and changed slowly during Phase II.

A short-duration peculiarity in the FP separation was observed at 20:48–20:49 UT, i.e. close to $t_2$, followed by the evident regained regular FP motion away from each other (see Figure 8a,b). This peculiarity may be explained by the visible disappearance of the more pronounced western FP (see panel *g* in Figure 3 in Liu et al. 2009). In turn, this disappearance can be caused by a short-duration change in the process of the magnetic reconnection or its location.

To complete our brief review of the image observations we refer to the paper of Liu et al. (2006). They presented a multi-wavelength study of the large-scale activities associated with this flare based on Hα, SXR, and *RHESSI* HXR data. Two brightening regions in Hα and SXR were located more than $2 \cdot 10^5$ km from the main flare site in the eastern and southern directions. These brightening regions were magnetically connected with the flare active region. The eastern kernel brightenings were observed till $t_2$ (see Figure 8d), while the southern ones occurred after $t_2$, i.e.

simultaneously with the third energy release. Liu et al. (2006) concluded that these remote brightenings, as well as all other activities, were due to the interaction of an erupting flux rope at the core of the flare with the magnetic field overlying the region.

Next, we compare the results of imaging observations considered above with the results of observations of HXR and gamma-ray emissions by the SONG. During Phase I fluxes of bremsstrahlung and gamma-lines observed by SONG increased (see Figures 1 and 6). The pion-decay emission was not detected during this phase and appeared only after 20:44 UT, i.e., with beginning of Phase II (see Figures 7d and 8e) which was manifested by the change of direction of FPs motion and by cessation of fast decrease of the flare shear. This emission flux grew up gradually till 20:48 UT and increased drastically at this time. The hardest spectrum of the parent protons was found at 20:48–20:51 UT. At the same time essential changes in the structure of the brightening regions occurred which were observed in Hα (see Figured 8d, and a certain feature in the FP motion was seen. Thus important changes occurred simultaneously both in particle acceleration and in a structure of the flare magnetic field. We believe that this simultaneity is not accidental but it is result of a common origin of these flare manifestations which are caused and controlled by the same process of continuing magnetic reconnection.

Apparently, the ion acceleration mechanisms or the sites of the acceleration are different when considerable changes in the flare structure occurred. In the above cited papers (Liu et al. (2006), Ji et al. (2008), Liu et al. (2009)), as well as in the papers of Krucker et al. (2005), Des Jardins et al. (2009), Xu et al. (2010), and Yang et al. (2011) continuing magnetic reconnection was associated with the erupting flux rope. These kinds of FP motions can be understood as sequential excitations/formations of flare loops (e.g., Somov et al. 2005; Schrijver et al. 2006; Schrijver 2009) when the primary reconnection site changes its location. This is consistent with magnetic reconnection occurring in the current sheet behind the CME, and confirms that the energy release processes for CMEs and flares are intimately related. It is worth recalling that Litvinenko and Somov (1995) demonstrated especially that a reconnecting current sheet forming behind the rising flux rope was a place where prolonged energy release occurred mainly in the form of high-energy particles. Based on this model we speculate: proton acceleration efficiency was growing up when the reconnecting current sheet arose, i.e. during Phases II. The most effective proton acceleration occurred near the time of the final formation of reconnecting current sheet, i.e., close to $t_2$.

The temporal correlation between the hard X-ray/microwave emission produced by accelerated particles and magnetic reconnection rate was first studied by Poletto and Kopp (1986), and then by Qiu et al. (2002, 2004). Li & Lin (2012) and Li et al. (2013) have also studied particle acceleration in the reconnection current sheet trailing the CME. A similar temporal correlation between the magnetic-flux change rate inferred by Miklenic et al. (2009) and the intensity of the pion-decay emission was observed in the 2003 October 28 flare (Kuznetsov et al. 2011). We also reported (Kurt & Yushkov 2014) an analogous association between the time evolution of accelerated proton spectrum hardness and the behavior of HXR/UV FP separation in the flares of 2001 August 25, 2003 October 28, and 2005 January 20 observed by CORONAS/SONG (paper in preparation). Although "post hoc, non est propter hoc" we consider that the process of particle acceleration is directly related to the change in the structure of the flare magnetic field.

## 6. SUMMARY

In this paper, we have mainly presented the results of the CORONAS/SONG observations of HXR and gamma-ray emissions from the 2003 October 29 solar flare in the 0.04–150 MeV energy range and their analysis.

1. These observations lasted from 20:38 to 20:58 UT and covered entirely the flare impulsive phase. An appreciable flux of the gamma-ray emission above 20 MeV was observed during this phase.

2. We have fitted count spectra measured by SONG using three-component model including the primary electron bremsstrahlung, prompt gamma-lines, and the pion-decay emission. Then we restored incident spectra of the flare emission. We also restored the flare emission spectra from the *RHESSI* data and found a reasonable agreement between SONG and *RHESSI* spectra in the 0.1–10 MeV energy range.

3. We have traced separately time profiles of components of HXR and gamma-rays and examined in this way evolution of different populations of accelerated particles. Electrons producing bremsstrahlung and ions with energies of 5–30 MeV producing prompt gamma-lines had a similar time history. There were two broad peaks at 20:42–20:47 and 20:50–20:54 UT.

4. We have argued that observed gamma-ray emission above 15 MeV had really the pion-decay origin. This fact proves that ions were accelerated to energies at least above the pion-production threshold (~300 MeV). The pion-decay emission appeared after 20:44 UT, grew sharply up to the maximum value at ~ 20:48 UT, and decreased after 20:51 UT.

5. We have estimated spectral indices of flare-accelerated protons assuming the power-law spectrum. We used fluxes of prompt gamma lines, $F_{4-7}$, and the pion-decay emission, $F_\pi$, derived from the SONG spectra as well as fluxes of the neutron-capture line, $F_{2.2}$, derived from the *RHESSI* spectra. The change of these indices during the course of the flare was traced. Spectral indices calculated from both $F_\pi/F_{4-7}$ and $F_{2.2}/F_{4-7}$ ratios coincided with regard to their errors, that implies the unified unbroken power-law spectrum in the proton energies range from 5 to several hundreds of MeV. The hardest ion spectrum, with a power-law index of $S = 3.4$–$3.6$, was found at 20:48−20:51 UT, i.e. during the maximum of the pion-decay emission.

      6. We compared the time history of the pion-decay emission with observed changes of the flare magnetic field structure derived mainly from *RHESSI* and Hα observations. We have found that at least two crucial features were seen simultaneously in these two series of observations. First, the pion-decay emission emerged after 20:44 UT, i.e. after general change of character of conjugated FP motion – from converging motion to motion away from each other. Second, an abrupt increase the pion-decay emission at 20:48 UT was close to essential change of structure of remote brightenings observed in Hα. We believe that this simultaneity is not accidental, but it is a result of a common origin of these manifestations of the flare. Both particle acceleration and changes of flare magnetic field structure are caused and controlled by the same process of continuing magnetic reconnection.

Acknowledgements.

The authors are grateful to G. Share for helpful discussions and for providing the gamma-line emission template; R. Murphy for providing the pion-decay emission spectra and to the *RHESSI* team for the excellent data. We are grateful to G. Fleishman for providing OVSA data. KK wishes to acknowledge the grant agency VEGA, project 2/0026/16. The work was supported by the Slovak Research and Development Agency under the contract number APVV-15-0194. This publication was supported by OP RDE, MEYS, Czech Republic under the project CRREAT, CZ.02.1.01/0.0/0.0/15_003/0000481. The authors are thankful to the anonymous Reviewer for his/her suggestions.

**Table 1**

Fitting results

| No. | UT | Fitting energy range, MeV | $\gamma$ | $E_0$, MeV | Flux, phot·cm$^{-2}$·s$^{-1}$·MeV$^{-1}$ | | |
|---|---|---|---|---|---|---|---|
| | | | | | 100 keV | 5 MeV | 100 MeV |
| 1 | 20:40:36–41:10 | 0.08–10.5 | 4.4 | 15 | $3.3 \cdot 10^2$ | $2.0 \cdot 10^{-2}$ | … |
| 2 | 20:41:10–42:22 | 0.08–10.5 | 3.8 | 5 | $5.9 \cdot 10^2$ | $5.1 \cdot 10^{-2}$ | … |
| 3 | 20:42:22–44:18 | 0.08–40 | 3.6 | 15 | $1.3 \cdot 10^3$ | $1.0 \cdot 10^{-1}$ | |
| 4 | 20:44:18–46:38 | 0.08–90 | 3.4 | 15 | $1.5 \cdot 10^3$ | $1.0 \cdot 10^{-1}$ | $(1.9 \pm 0.8) \cdot 10^{-5}$ |
| 5 | 20:46:38–47:51 | 0.08–150 | 3.4 | 25 | $8.0 \cdot 10^2$ | $6.0 \cdot 10^{-2}$ | $(9.3 \pm 4.9) \cdot 10^{-5}$ |
| 6 | 20:47:51–49:44 | 0.08–150 | 3.2 | 15 | $3.8 \cdot 10^2$ | $5.1 \cdot 10^{-2}$ | $(2.3 \pm 0.5) \cdot 10^{-4}$ |
| 7 | 20:49:44–51:04 | 0.08–150 | 3.4 | 20 | $7.7 \cdot 10^2$ | $7.8 \cdot 10^{-2}$ | $(2.6 \pm 0.6) \cdot 10^{-4}$ |
| 8 | 20:51:04–52:08 | 0.08–60 | 3.2 | 8 | $9.6 \cdot 10^2$ | $9.9 \cdot 10^{-2}$ | $(7.6 \pm 5.4) \cdot 10^{-5}$ |
| 9 | 20:52:08–53:00 | 0.08–60 | 3.2 | 15 | $6.1 \cdot 10^2$ | $4.5 \cdot 10^{-2}$ | $(1.1 \pm 0.8) \cdot 10^{-4}$ |
| 10 | 20:53:00–54:00 | 0.08–10.5 | 3.0 | 20 | $4.3 \cdot 10^2$ | $4.6 \cdot 10^{-2}$ | … |
| 11 | 20:54:00–55:00 | 0.08–150 | 3.0 | 20 | $4.0 \cdot 10^2$ | $4.8 \cdot 10^{-2}$ | $(5.5 \pm 5.5) \cdot 10^{-5}$ |
| 12 | 20:55:00–56:00 | 0.08–22.5 | 3.0 | 20 | $3.4 \cdot 10^2$ | $4.2 \cdot 10^{-2}$ | … |
| 13 | 20:56:00–57:00 | 0.08–22.5 | 2.8 | 15 | $2.0 \cdot 10^2$ | $3.6 \cdot 10^{-2}$ | … |

**Table 2**

HXR and gamma-ray component fluxes and parameters of parent proton spectra

| No | UT | Flux, phot·cm$^{-2}$·s$^{-1}$ | | | $R$ | $S$ |
|---|---|---|---|---|---|---|
| | | Bremsstr. >100 keV | γ-lines, 4–7 MeV | π-decay emission, 30–100 MeV | | |
| 1 | 20:40:36–20:41:10 | $1.4 \cdot 10^1$ | $6.4 \cdot 10^{-2}$ | … | … | … |
| 2 | 20:41:10–20:42:22 | $2.9 \cdot 10^1$ | $1.7 \cdot 10^{-1}$ | … | … | … |
| 3 | 20:42:22–20:44:18 | $6.7 \cdot 10^1$ | $3.2 \cdot 10^{-1}$ | … | … | … |
| 4 | 20:44:18–20:46:35 | $7.9 \cdot 10^1$ | $3.2 \cdot 10^{-1}$ | $(2.1 \pm 5.0) \cdot 10^{-3}$ | $(6.7 \pm 14.0) \cdot 10^{-3}$ | $4.3^{+3.0}_{-0.3}$ |
| 5 | 20:46:35–20:47:51 | $4.3 \cdot 10^1$ | $1.9 \cdot 10^{-1}$ | $(1.1 \pm 0.7) \cdot 10^{-2}$ | $(5.6 \pm 3.8) \cdot 10^{-2}$ | $3.8^{+0.3}_{-0.1}$ |
| 6 | 20:47:51–20:49:44 | $2.2 \cdot 10^1$ | $1.5 \cdot 10^{-1}$ | $(2.6 \pm 0.6) \cdot 10^{-2}$ | $(1.7 \pm 0.4) \cdot 10^{-1}$ | $3.5 \pm 0.1$ |
| 7 | 20:49:44–20:51:04 | $4.2 \cdot 10^1$ | $2.4 \cdot 10^{-1}$ | $(2.9 \pm 0.7) \cdot 10^{-2}$ | $(1.2 \pm 0.3) \cdot 10^{-1}$ | $3.6 \pm 0.1$ |
| 8 | 20:51:04–20:52:08 | $5.6 \cdot 10^1$ | $3.1 \cdot 10^{-1}$ | $(8.7 \pm 6.3) \cdot 10^{-3}$ | $(2.8 \pm 2.0) \cdot 10^{-2}$ | $4.0^{+0.3}_{-0.2}$ |
| 9 | 20:52:08–20:53:00 | $3.5 \cdot 10^1$ | $1.4 \cdot 10^{-1}$ | $(1.2 \pm 0.5) \cdot 10^{-2}$ | $(9.0 \pm 6.7) \cdot 10^{-2}$ | $3.7 \pm 0.1$ |
| 10 | 20:53:00–20:54:00 | $2.7 \cdot 10^1$ | $1.4 \cdot 10^{-1}$ | … | … | … |
| 11 | 20:54:00–20:55:00 | $2.3 \cdot 10^1$ | $1.5 \cdot 10^{-1}$ | $(6.3 \pm 6.3) \cdot 10^{-3}$ | $(4.2 \pm 4.2) \cdot 10^{-2}$ | $3.9^{+3.0}_{-0.2}$ |
| 12 | 20:55:00–20:56:00 | $2.1 \cdot 10^1$ | $1.3 \cdot 10^{-1}$ | … | … | … |
| 13 | 20:56:00–20:57:00 | $1.4 \cdot 10^1$ | $1.1 \cdot 10^{-1}$ | … | … | … |

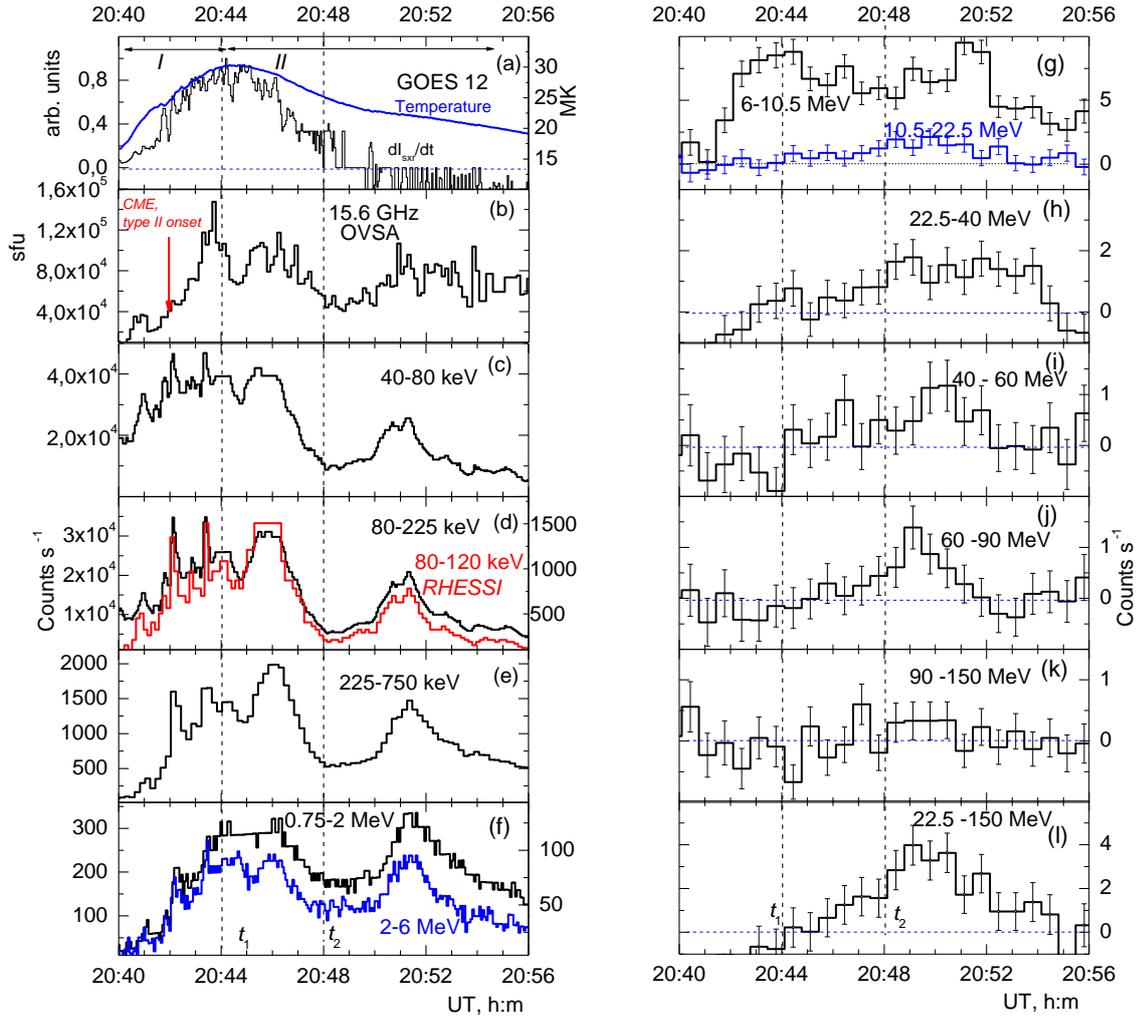

**Figure 1**. Multi-wavelength time history of the 2003 October 29 solar flare. (a) The GOES-12 SXR derivative, $dI_{SXR}/dt$, and the SXR-source temperature; (b) - microwave flux (15.6 GHz, OVSA). Red arrow marks onset time of CME and type II radio-emission (from Liu et al. 2006); (c-l) the SONG net count rates (black and blue curves). The *RHESSI* HXR data are added in panel d (red curve, right Y-axes, from Liu et al. 2009). Division to Phases I and II and transition times, $t_1$ and $t_2$, are taken from Liu et al. (2009).

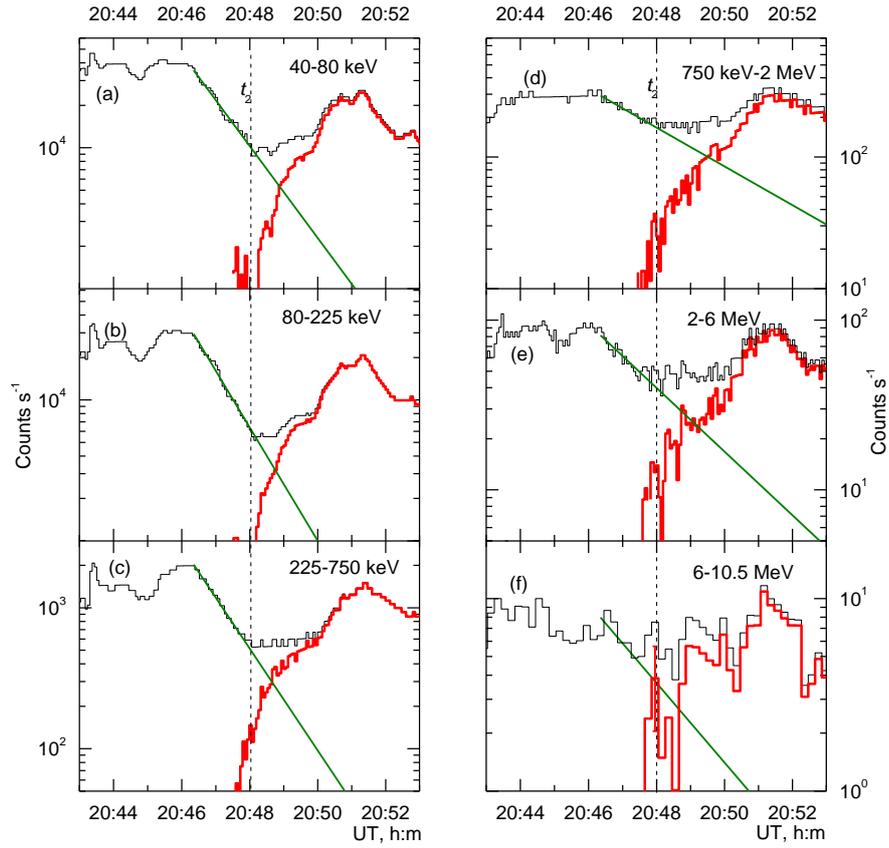

**Figure 2.** Net count rates recorded by SONG in the low-energy channels – black curves. The *e*-folding decay in each channel is shown by olive line. Time profiles of the count rate derived after this decay subtraction are shown by red lines.

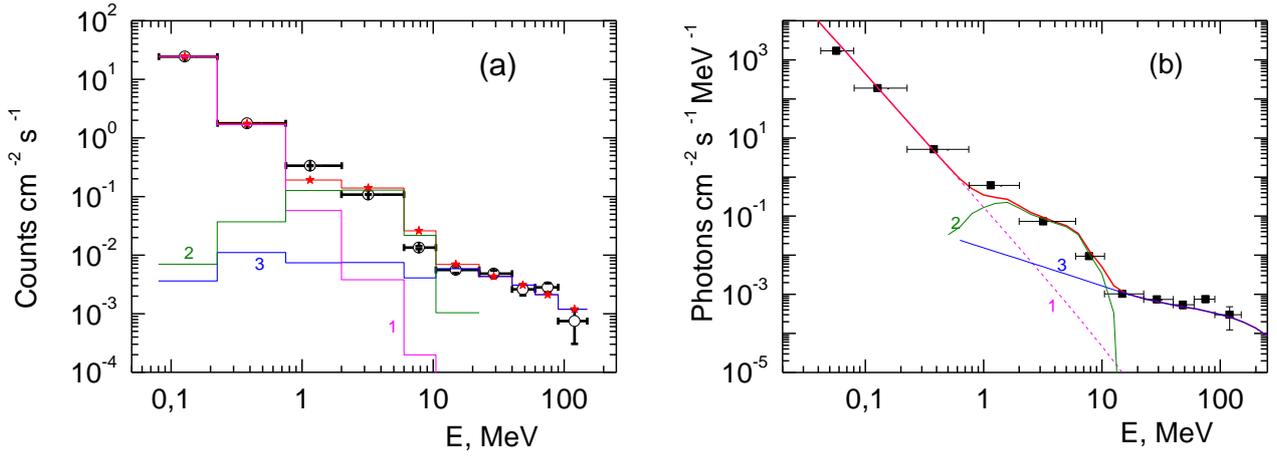

**Figure 3.** (a) Background-subtracted SONG count spectrum accumulated between 20:47:53 and 20:51:00 UT (black points with open circles). This spectrum has been fitted by a power law with exponential cutoff (magenta histogram *1*), nuclear de-excitation component (green *2*), and a pion-decay component (blue *3*). The total fitted spectrum is shown by red histogram with stars. (b) Comparison of the same observed count spectrum recalculated to the photon one with the restored gamma-ray spectrum. Numbers and colors are the same as in panel (a).

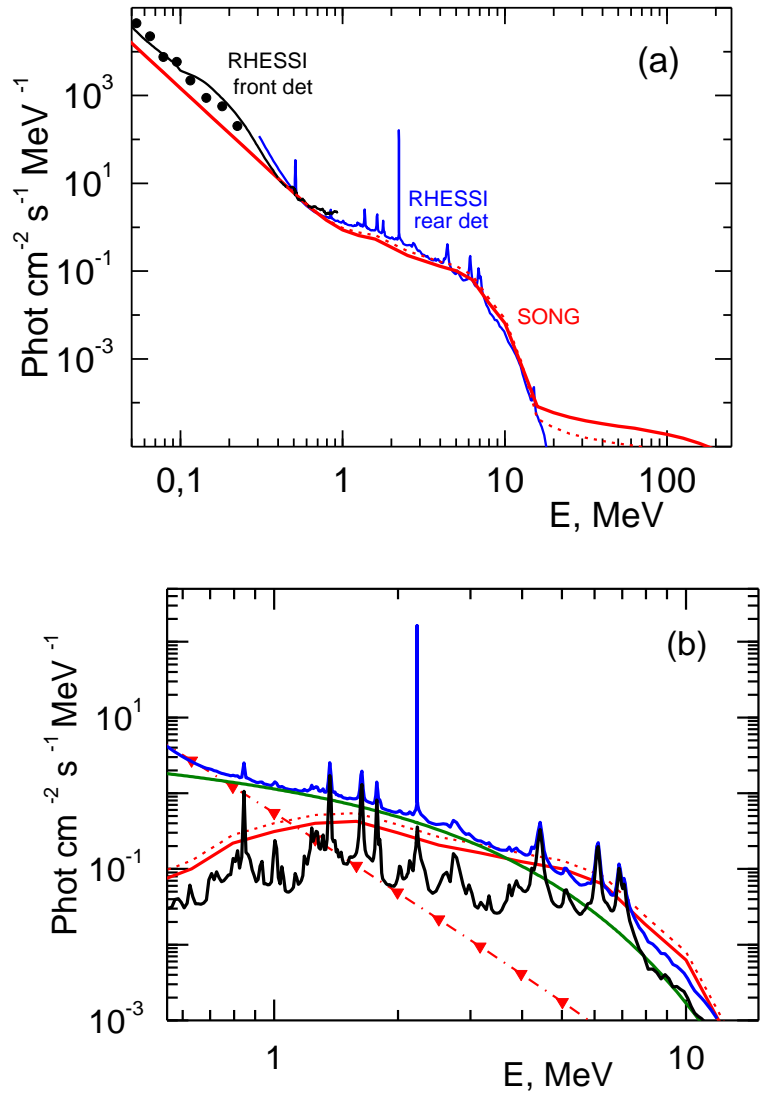

**Figure 4** (a) HXR and gamma-emission spectrum from the 2003 October 29 flare at 20:44:18–20:46:35 UT. The spectrum derived from total SONG count spectra is shown by the dashed red curve, the spectrum without 2.223 MeV line deposit in the SONG count rates – by solid red curve. The spectra measured by *RHESSI* front detectors are shown by black curves; spectra from rear detectors are shown by blue curves, and spectra derived from images – by black circles. (b) Enlarged part of the same spectrum. The total spectrum obtained from the rear *RHESSI* detectors is shown by blue curve, its components are shown by green (PL2) and black (gamma-lines) curves. Dashed red curve presents the gamma-line component obtained from the SONG data before subtraction of the 2.223 MeV line contribution, solid red line - the same component obtained after this subtraction. Dashed-dotted red line with triangles presents the bremsstrahlung spectrum observed with SONG.

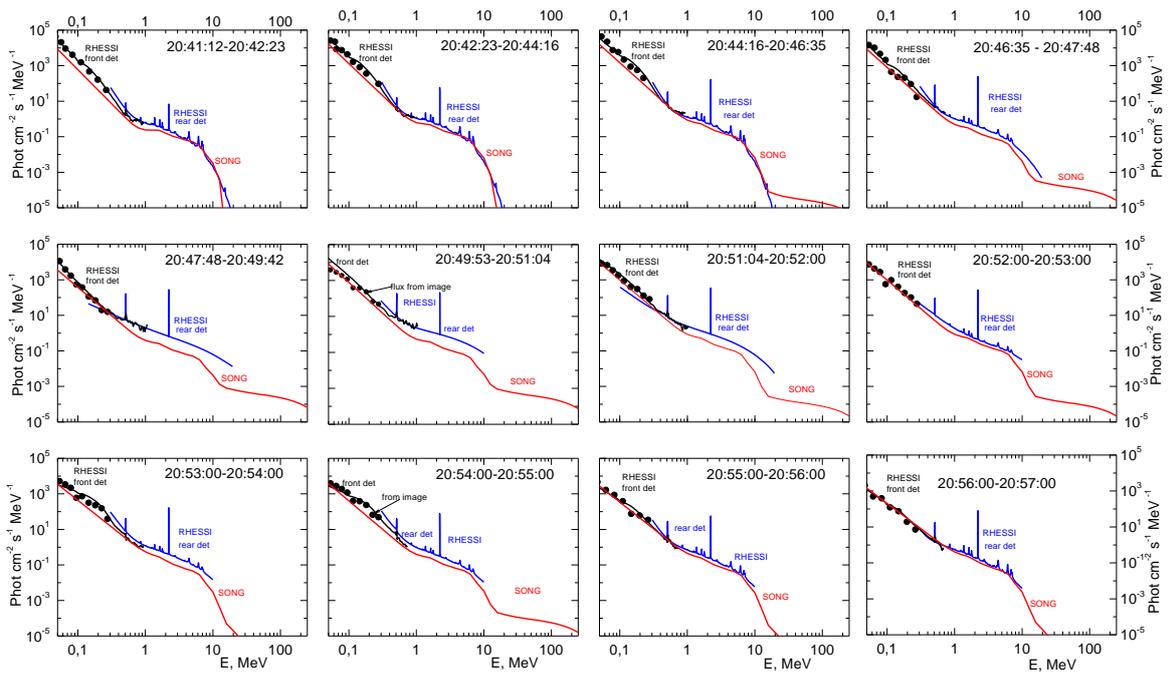

**Figure 5.** HXR and gamma-ray spectra from the 2003 October 29 flare. The spectra obtained from the SONG data are shown by red curves. The spectra obtained from the *RHESSI* front detectors are shown by black curves, from the rear detectors by blue curves. Spectra obtained from images are shown by black circles.

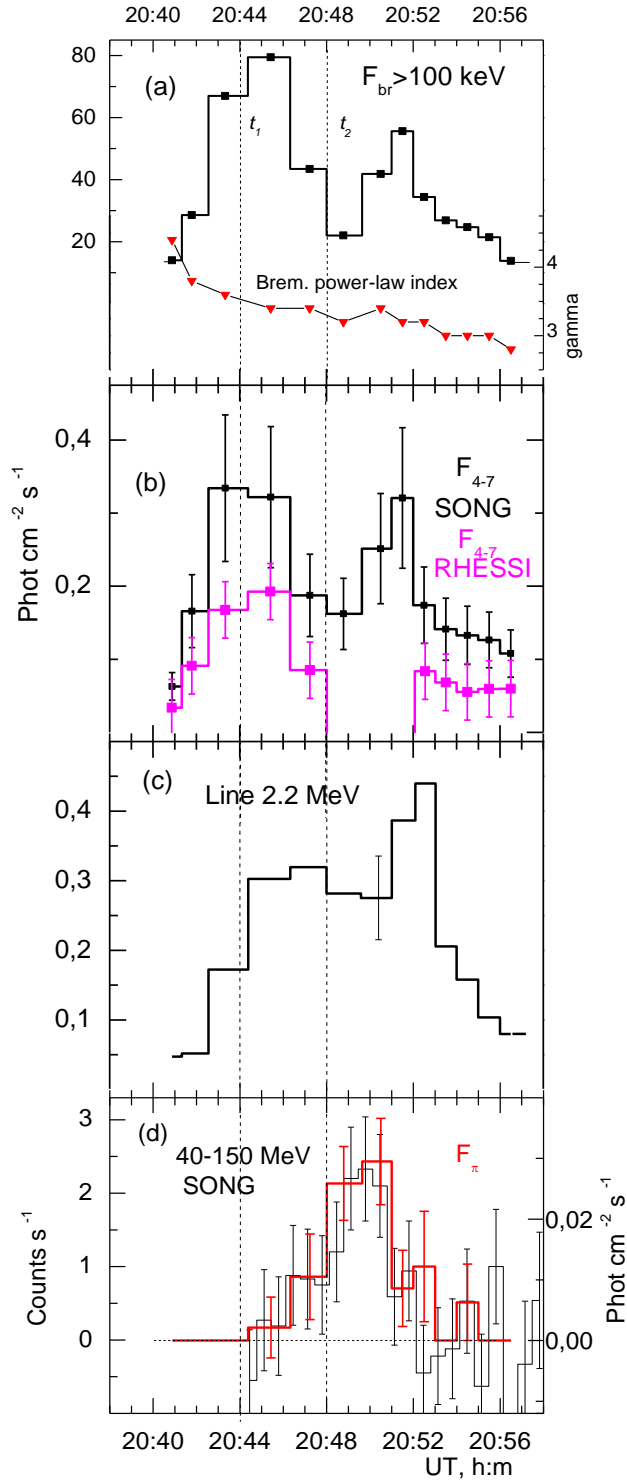

**Figure 6.** (a) The bremsstrahlung flux (black curve) and its power-law index (red curve with triangles), measured by SONG. (b) Fluxes of gamma-lines in the 4–7 MeV range measured by SONG (black) and *RHESSI* (magenta). (c) Flux of the 2.223 MeV line measured by *RHESSI*. (d) Flux of the pion-decay emission (red) and the SONG count rates (black).

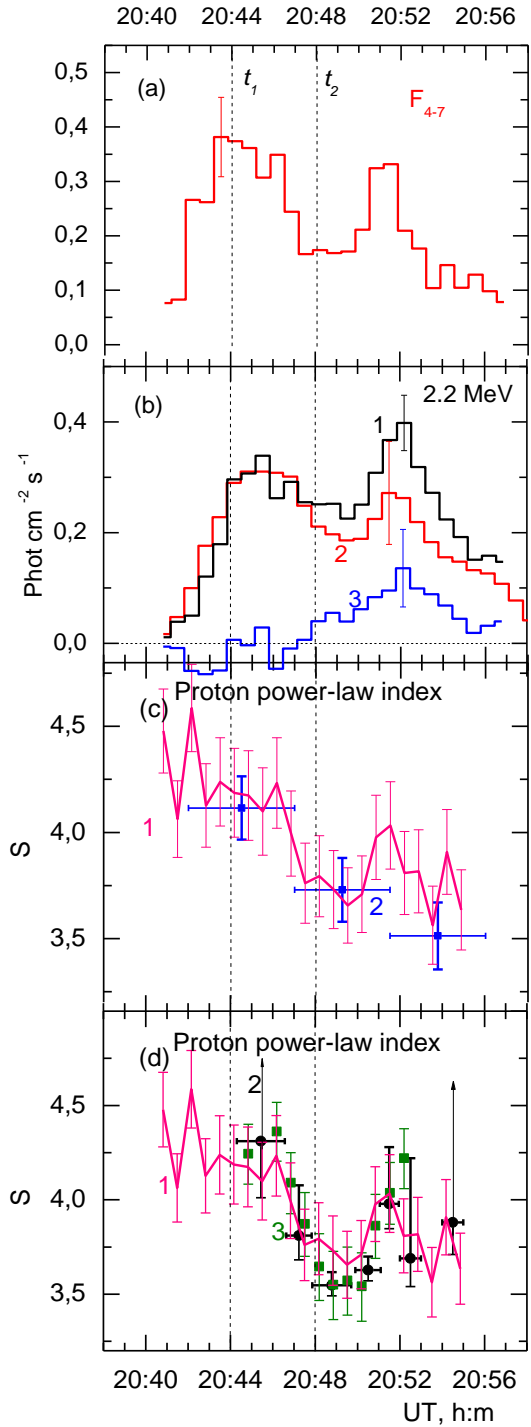

**Figure 7.** (a) Flux of gamma-lines in the 4–7 MeV range measured by SONG. (b) Flux of the 2.223 MeV line measured by *RHESSI* – *1* (black); simulated flux of the 2.223 MeV line – *2* (red); residuals between them – *3* (blue). (c) Proton power-law index obtained from the $F_{2.2}/F_{4-7}$ ratio for 40-s (*1*) and more long intervals (*2*). (d) Proton power-law index obtained: *1*– from the $F_{2.2}/F_{4-7}$ ratio; *2*– from the $F_{\pi}/F_{4-7}$ ratio; *3* - the same as *2*, but calculated for 40-s intervals.

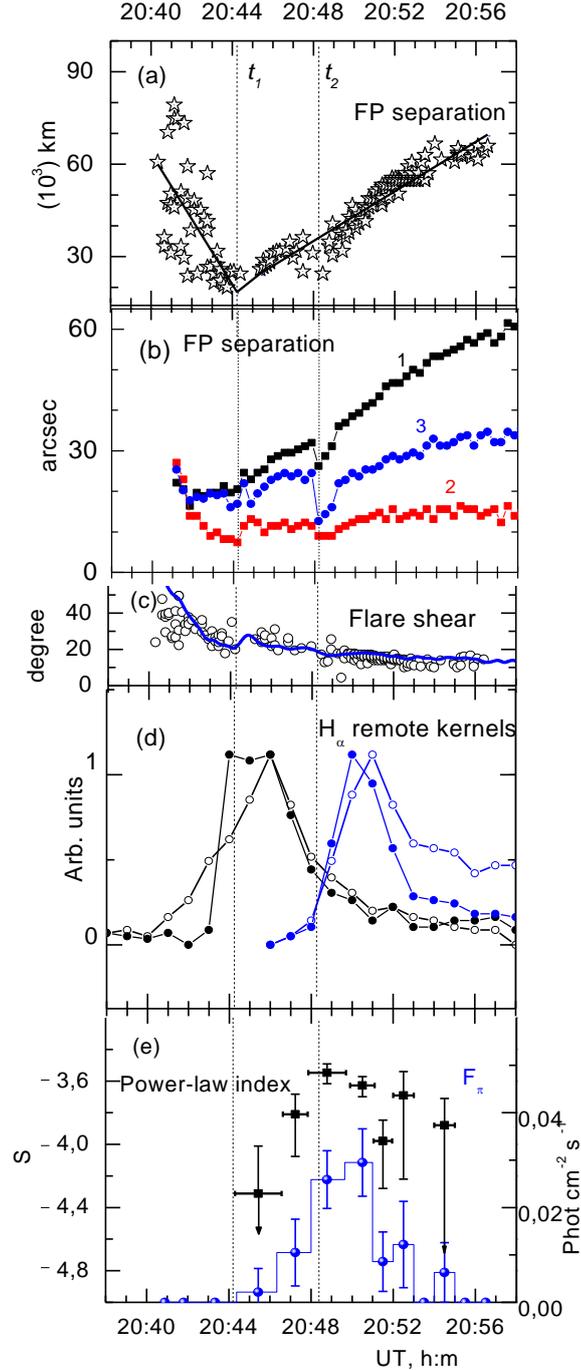

**Figure 8.** (a) The distance between conjugated FPs (from Ji et al. 2008); (b) orthogonal components of the separation between western and eastern FPs at 60–100 keV (from Liu et al. 2009): 1 – perpendicular, and 2 – parallel to the NL; 3 – perpendicular distance to the NL from the western FP; (c) flare shear; black points are taken from Ji et al. (2008), blue curve - from Liu et al. (2009); (d) intensities of the major remote Hα brightening patches in the eastern (black curves) and southern (blue curves) sites (from Liu et al. 2006). Filled and unfilled circles correspond to different kernels; (e) Flux of the pion-decay emission (blue points) and proton power-law spectrum index obtained from the $F_\pi/F_{4-7}$ ratio (black points).